\documentclass[sigconf,authorversion,nonacm]{acmart}
\AtBeginDocument{%
  \providecommand\BibTeX{{%
    \normalfont B\kern-0.5em{\scshape i\kern-0.25em b}\kern-0.8em\TeX}}}

\usepackage{subcaption}

\catcode`\"=13     
\def"#1"{``#1''}   

\usepackage{bbold}

\theoremstyle{definition}
\newtheorem{definition}{Definition}
\newtheorem{proposition}[definition]{Proposition}

\acmISBN{978-1-4503-XXXX-X/18/06}

\begin{document}

\title[Rethinking Algorithmic Fairness from the Perspective of Approximate Justice]{What's Distributive Justice Got to Do with It? Rethinking Algorithmic Fairness from the Perspective of Approximate Justice}


\author{Corinna Hertweck}
\affiliation{%
  \institution{University of Zurich \& Zurich University of Applied Sciences}
  \city{Zurich}
  \country{Switzerland}
}
\email{corinna.hertweck@zhaw.ch}

\author{Christoph Heitz}
\affiliation{%
  \institution{Zurich University of Applied Sciences}
  \city{Zurich}
  \country{Switzerland}
}
\email{christoph.heitz@zhaw.ch}

\author{Michele Loi}
\affiliation{%
  \institution{AlgorithmWatch}
  \city{Berlin}
  \country{Germany}
}
\email{loi@algorithmwatch.org}

\renewcommand{\shortauthors}{Anonymous Authors}

\begin{abstract}
  In the field of algorithmic fairness, many fairness criteria have been proposed. Oftentimes, their proposal is only accompanied by a loose link to ideas from moral philosophy -- which makes it difficult to understand when the proposed criteria should be used to evaluate the fairness of a decision-making system. More recently, researchers have thus retroactively tried to tie existing fairness criteria to philosophical concepts. Group fairness criteria have typically been linked to egalitarianism, a theory of distributive justice.
  This makes it tempting to believe that fairness criteria mathematically represent ideals of distributive justice and this is indeed how they are typically portrayed. In this paper, we will discuss why the current approach of linking algorithmic fairness and distributive justice is too simplistic and, hence, insufficient. We argue that in the context of imperfect decision-making systems -- which is what we deal with in algorithmic fairness -- we should not only care about what the ideal distribution of benefits/harms among individuals would look like but also about how deviations from said ideal are distributed. Our claim is that algorithmic fairness is concerned with unfairness in these deviations.
  This requires us to rethink the way in which we, as algorithmic fairness researchers, view distributive justice and use fairness criteria.
\end{abstract}



\received{20 February 2007}
\received[revised]{12 March 2009}
\received[accepted]{5 June 2009}

\maketitle

\section{Introduction}\label{sec:introduction}

In the algorithmic fairness literature, there are numerous fairness criteria and metrics that try to operationalize the complex and context-dependent construct that is fairness~\cite{jacobs2021measurement}. As these criteria and metrics have often been proposed without a deeper engagement with the philosophical literature, there is a wave of work that has tried to tie the proposed statistical metrics of fairness to the philosophical debate. This is particularly the case for group fairness criteria and egalitarianism (see, e.g., \cite{heidari2019moral,binns2018fairness,binns2020apparent,gajane2017formalizing}).
Egalitarianism is a theory of distributive justice and, as such, tells us something about how benefits/harms should be distributed among members of a society. There is thus a parallel to algorithmic decision-making systems that make decisions about individuals that result in different benefits/harms and could therefore be seen as distributing benefits/harms in a population. The fact that fairness criteria for decision-making systems have been linked to distributive justice thus seems intuitive.
However, we believe that this often-made connection is blurred and leaves an important conceptual question unanswered: \textit{If theories of distributive justice are defined at the level of individuals, how do they relate to group fairness criteria defined at the level of socio-demographic groups?}
In this paper, we will rethink the concept of distributive justice and how it relates to algorithmic fairness. Based on this, we present our understanding of what we think is usually meant by "algorithmic fairness".
The crux is this: Patterns of distributive justice like egalitarianism, sufficientarianism or maximin cannot be perfectly fulfilled in practice. If a pattern was perfectly fulfilled (e.g., for egalitarianism: all individuals receive the same outcome), then there would be no structural injustice (in how the pattern is fulfilled). If, however, the pattern is not perfectly fulfilled – which is the case with any real-world system as there will always be deviations from the ideal – then individual deviations from the ideal distribution (i.e., when an individual does not receive what they should receive under the ideal distribution\footnote{There could be multiple ideal distributions. There are, for example, multiple possible distributions that fulfill egalitarianism. One then has to decide how to measure the deviation from the ideal by deciding on one ideal. For egalitarianism, one option would be to measure the deviations from the distribution where everyone gets the mean of what everyone gets under the current distribution.}) might affect some groups more than others – in a biased way. This is what we will call \textit{structural injustices}. Generally speaking, structural injustices are, for example, unfair opportunities (which are unjust according to \citet{rawls1999theory} fair equality of opportunity principle) or systematic, structural discrimination against a certain category of people, which is also clearly unjust.%
\footnote{Note that \citet{kasirzadeh2022algorithmic} already points out the relationship between algorithmic fairness and structural injustice. She defines the latter based on the work of Iris Marion Young. However, she finds that "algorithmic fairness does not accommodate structural injustices in its current scope" \cite[p. 349]{kasirzadeh2022algorithmic} because mathematical fairness criteria do not consider the sociotechnical nature of automated decision-making systems and instead just condense fairness into a single metric that does not consider "power structures and social dynamics" \cite[p. 351]{kasirzadeh2022algorithmic}. We strongly agree with this view and urge practitioners not to misinterpret the fulfillment of statistical criteria as proof of justice or fairness. However, our paper will interpret the term "structural injustice" only in the context of what we can evaluate given the data of a decision-making system (and we will therefore sidestep questions of, e.g., procedural fairness).} In the terminology of distributions, we measure structural injustices as unfairly distributed individual deviations from an ideal distribution. We argue that this is what algorithmic fairness metrics are supposed to capture even though the algorithmic fairness literature has not made this clear so far. Note that this is \textit{our} interpretation of the fairness concern in the algorithmic fairness literature because there is no agreed-upon definition of what we mean by fairness in algorithmic fairness. Our approach is different from others as algorithmic fairness is typically associated with theories of distributive justice themselves. We, however, argue that algorithmic fairness metrics do \textbf{not} check whether a theory of distributive justice is fulfilled, and instead, it is the deviation from said theory that fairness is concerned with. We thus rethink every theory of distributive justice as follows:\footnote{Note that we will assume that they are defined at the individual level, which is the case for the majority of them. An exception to this is Rawls's theory of justice \cite{rawls1999theory}} There is a distributive pattern that defines how benefits / harms should be distributed between individuals. If this is perfectly fulfilled, then this is where the evaluation of the fulfillment of the theory of distributive justice ends. However, if there is a system where mistakes will be made, then it is not enough to test for the pattern of justice and we also need to test for structural injustice.
We will clarify this conceptual map with a concrete discussion of what this means for distributions that are supposed to follow \textit{egalitarianism}, \textit{sufficientarianism} or \textit{maximin}.

In Section~\ref{sec:retracing}, we will retrace how algorithmic fairness, and more specifically, group fairness, has been connected to theories of distributive justice thus far. Based on the issues we find in this connection, we present our proposal for how to map algorithmic fairness to distributive justice in Section~\ref{sec:deconstruction}. This contains an explanation of how we view distributive justice and its relation to structural injustices. Next, we apply this mapping in Section~\ref{sec:individual-group} to demonstrate what this means for algorithmic fairness criteria in the context of egalitarianism, sufficientarianism or maximin. Section~\ref{sec:naive-operationalizations} highlights the necessity of our conceptual mapping by comparing our resulting fairness criteria from Section~\ref{sec:individual-group} to more naive fairness criteria that are based on the current status of the debate. Section~\ref{sec:mittelstadt} discusses how the work of \citet{mittelstadt2023unfairness} fits into our conceptual map. Finally, we discuss the implications of our work and possible future directions of research in Section~\ref{sec:conclusion}.

\section{Retracing the Debate: Distributive Justice and Group Fairness Criteria}\label{sec:retracing}

The parallels between distributive justice and automated decision-making systems mentioned in Section~\ref{sec:introduction} have led to the algorithmic fairness literature drawing parallels between distributive justice and algorithmic fairness. This section will retrace this development after introducing the basics of both concepts.

\subsection{Theories of Distributive Justice}

Theories of distributive justice propose how benefits and burdens should be distributed in society~\cite{sep-distributive-justice}. They can vary in whether they care about the just distribution between individuals or groups~\cite{sep-distributive-justice}. However, theories of distributive justice are typically concerned with individuals~\cite{sen1980equality}. One notable exception from this is Rawls' second principle, which states that “social and economic inequalities are to be arranged so that they are both (a) reasonably expected to be to everyone’s advantage, and (b) attached to positions and offices open to all” \cite[p. 53]{rawls1999theory}.
Criterion (a) is known as the difference principle. All illustrations of the difference principle are given by considering the average expectations of different “social classes”~\cite[p. 67]{rawls1999theory} and what Rawls refers to as the “representative man [sic.] who is worst off”~\cite[p. 68]{rawls1999theory}. This representative person is a clear statistical or sociological generalization.
However, the subsequent discussion within analytic philosophy focuses on a more abstract conception of justice disconnected from sociological categories. This is important to note as this makes it difficult to think about structural injustice at the group level for such theories.

\subsubsection{Egalitarianism and Beyond}

\textbf{Egalitarianism} is a theory of distributive justice that sees equality per se as morally desirable. According to egalitarianism, people should thus receive the same outcomes, be treated in the same way, be seen in the same way or be equal in some other way. Egalitarianism can thus take different forms depending on what one sees as the metric that ought to be equal.\footnote{Note that egalitarians advocate for equality in a qualified form. A common form of egalitarianism is, for example, “luck egalitarianism”: Both \citet{cohen1990equality} and \citet{temkin2017equality} defend this idea of equality (even at the cost of leveling down) as the best description of what justice is (before compromise with other values). What they mean by equality is  “equality in conditions that are not the result of choice”. A plausible alternative is to say that equality is an appealing idea when it is framed in terms of equal desert. This is the “desert egalitarian” form. \citet{shelly2012ratio} frames “desert” as moral desert. Alternatively, desert can also be a placeholder for any other justification of inequality~\cite{loi2021fair}, e.g., when used in generic sentences such as “people with more urgent needs deserve more urgent assistance by the state”.}
While egalitarianism is intuitive, it has one remarkably unpalatable implication. Egalitarianism between two people can be achieved not only by increasing the benefit of one person but also by lowering the benefit of the other person (or even by lowering the benefit of both to a lower level). Thus, egalitarianism can be achieved by worsening the results for both people, which is often referred to as the ``leveling down'' objection.
A reasonable agent may reflect that there are contexts in which achieving equality by leveling down would be considered a worsening from the point of view of fairness by most affected people. It would then seem unreasonable to impose fairness that may worsen the prospects of all the groups involved – especially when the affected groups themselves refuse it.

Thus, if one cares about the social acceptability of fair prediction-based decisions, one should either abandon egalitarianism or suitably modify it in order to align it with stakeholders' view of fairness – at least to the extent that these appear to be reasonable.
The most obvious way to do this is suggested by the history of the debate on the nature of justice in analytic political philosophy. Analytic political philosophers have been debating for decades the nature of fairness. Since the seminal work of John Rawls~\cite{rawls1999theory} alternatives to simple equality have emerged, such as the \textbf{difference principle}, that requires maximizing the resources of the least advantaged group in absolute terms. Note that the difference principle is the application of maxmin to a particular context and given specific boundary conditions. Hence, we refer to \textbf{maxmin} in the context of various operationalizations (where the context differs, sometimes significantly sometimes in subtle ways) from the one of Rawls’s difference principle.
\textbf{Sufficientarianism} also does not recognize equality's intrinsic value: according to sufficientarianism, equality simply has no moral value once all people's needs are satisfied, and it only has value to the extent that it is instrumental in meeting people’s needs, or some other threshold of well-being that is equally significant for a human person as having a fundamental need fulfilled (e.g., a "non-basic" need, such as personal realization)~\cite{frankfurt2018equality}. This alternative becomes particularly persuasive when the quantity of a limited resource deemed sufficient aligns with the quantity necessary to prevent significant harm or catastrophe. Consider, for example, a scenario where there is a limited supply of a particular resource, such as food or medicine, which is sufficient to sustain some but not all individuals within a population. Let us assume the population consists of ten members, each requiring a minimum of five units of the resource to survive, with a total of forty units available. In such a case, for any members of the population to survive, some must receive a larger share than others. Distributing the resource equally, providing each person with four units, results in the direst outcome: the demise of everyone.~\cite{frankfurt2018equality} Yet other contexts might suggest other normative goals, such as prioritarianism, which we will, however, not discuss in this paper.

These are situations where the particular context makes equality an unattractive normative goal that should be replaced by another pattern. Contextual pluralism seems to be necessary: According to a contextually pluralist approach, different patterns of justice may be most reasonable depending on features of the context, of the subjects affected, and the agents involved.

\subsubsection{Criterion-Type and Optimization-Type Theories of Distributive Justice}

We propose a categorization of theories of distributive justice into \textit{criterion-type} and \textit{optimization-type} theories. Criterion-type theories of distributive justice have a well-defined normative goal, for which we can check whether it is fulfilled. These types of theories have counterfactual-invariant goals: We can always say whether either is achieved by observing the actual distribution and it is entirely irrelevant what other distributions could be realized. This is the case for egalitarianism and sufficientarianism. By contrast, maxmin requires us to compare the actual distribution to counterfactual ones (e.g., feasible alternatives) as its goal is the optimization of a value. How well we do can only be understood in comparison to possible alternatives. We will refer to this as optimization-type theories of distributive justice.\footnote{Other examples of this are prioritarianism and utilitarianism.} For the operationalization of theories of distributive justice, which we will get to later, it is important to understand these types.

\subsection{Group Fairness Criteria}\label{sec:algorithmic-fairness}

Standard group fairness \textit{criteria} such as statistical parity ($P(D=1|A=a)=P(D=1|A=b)$), equality of opportunity ($P(D=1|Y=1, A=a)=P(D=1|Y=1, A=b)$)~\cite{hardt2016equality} or positive predictive value parity ($P(Y=1|D=1, A=a)=P(Y=1|D=1, A=b)$) demand equality in some metric across socio-demographic groups. What changes between them is \textit{what} should be equal and for \textit{whom} this should be equal. Statistical parity, for example, demands equal rates of positive decisions across groups while equality of opportunity also demands that but only for the parts of the group with a positive $Y$ label.
If we assume a \textit{measure} $M$ that is measured at the group level, these fairness criteria thus take the form $M(A=a)=M(A=b)$.
In this paper, we take a utility-based approach (as previously suggested by~\cite{binns2018fairness, finocchiaro2021bridging, jorgensen2023not, ben2021protecting}), meaning that instead of comparing shares of positive decisions, we assume that we want to compare their consequences as measured by the utility $U$ of a decision. We thus assume continuous utility values and compare these continuous values between groups. We take this approach for simplicity's sake to illustrate a more general point than it would have been possible with a binary utility (which is what is often assumed in the algorithmic fairness literature).

\subsection{Status of Conceptual Connection and Open Questions}

Early work proposed group fairness criteria based on standard performance metrics such as accuracy and measures of the confusion matrix (see, e.g., \cite{pedreshi2008discrimination, hardt2016equality, verma2018fairness, corbett2023measure}). Starting in 2016, the infamous COMPAS case~\cite{machine_bias} started a debate about the compatibility of fairness criteria (see \cite{kleinberg2016inherent,chouldechova2017fair}). With that came the more urgent need for an understanding of fairness criteria from a philosophical and normative perspective. Attempts to connect group fairness criteria to philosophical theories mostly connected them to forms of egalitarianism. \citet{gajane2017formalizing} and \citet{heidari2019moral}, for example, connect different group fairness criteria to different forms of egalitarianism while \citet{binns2018fairness} gives a philosophical account of discrimination and egalitarianism to show "that ‘fairness’ as used in the fair machine learning community is best understood as a placeholder term for a variety of normative egalitarian considerations" \cite[p. 2]{binns2018fairness}.
More recently, \citet{kuppler2021distributive} has criticized algorithmic fairness's focus on egalitarianism and discussed other theories of distributive justice that the literature could make use of. Some of these have already been implemented -- also as group fairness criteria (see, e.g., \cite{martinez2020minimax, diana2021minimax}).
However, in a lot of this work, the link between algorithmic fairness and distributive justice remains vague and underspecified. \cite{mittelstadt2023unfairness}, for example, writes that "Group fairness measures are related to egalitarian thinking in distributive justice" \cite[p. 19]{mittelstadt2023unfairness}. It seems intuitive to connect distributive justice to algorithmic fairness as both are about the distribution of benefits/harms, but an important question remains open: \textit{Theories of distributive justice typically operate at the individual level, but how do they relate to group fairness criteria that operate at the level of socio-demographic groups?}
We claim that this unclarity is based on a fundamental misconception about how algorithmic fairness and distributive justice relate. We address this point with our rethinking of distributive justice.

\section{Rethinking Algorithmic Fairness as Approximate Justice}\label{sec:deconstruction}

In this section, we present how we view distributive justice in relation to algorithmic fairness. We will also introduce the terminology that we will use in this context, such as distributive justice and structural injustice.

\subsection{Differentiating Theories of Distributive Justice, Structural Injustices and Distributive Patterns}

Broadly speaking, our rethinking of distributive justice in the context of algorithmic decision-making builds on the hypothesis that decision-making systems are never perfect and that an ideal distribution cannot be reached in the real world, so we have to allow for deviations from perfect distributive justice.
However, when we allow for deviations from the ideal, there is a chance that these individual deviations are unfairly distributed. It could be that a group is systematically more likely to be a victim of disadvantageous deviations. This is what we would then consider to be a \textit{structural injustice} and what -- so we argue -- we want to test for in algorithmic fairness.
This can be done with fairness criteria that operate at an individual level (e.g., through individual fairness~\cite{dwork2012fairness} as pointed out by~\cite{binns2020apparent} or through causal models~\cite{kusner2017counterfactual}), but structural injustice is difficult to evaluate at the individual level. This is where socio-demographic groups come in as they have the practical advantage that we can more easily look for patterns of structural injustices -- this is the basis of group fairness criteria.

In theory, there are thus two options when we want to evaluate whether a theory of distributive justice is fulfilled:
\begin{enumerate}
    \item Chosen theory of distributive justice is perfectly fulfilled (measured at the level of individuals) $\rightarrow$ no further tests necessary
    \item Chosen theory of distributive justice is not perfectly fulfilled $\rightarrow$ (a) check if the theory of distributive justice is approximately fulfilled (measured at the level of individuals) and (b) check for signs of structural injustices (group fairness criteria measure this at the level of socio-demographic groups)
\end{enumerate}

Thus, whenever our chosen theory of distributive justice cannot be perfectly fulfilled (which is essentially always the case in the real world),\footnote{Note that egalitarianism, sufficientarianism and maximin are what \citet{rawls1999theory} calls "ideal theories" that assume ideal conditions, such as that every individual follows the rules. As \citet{fazelpour2020algorithmic} criticize, algorithmic fairness focuses on ideal theories of justice instead of non-ideal ones (which are needed in the real world).} we have to check for structural injustices. One might ask why most theories of distributive justice do not account for this. Presumably, this is because most theories are not concerned with their operationalization. They instead tend to focus on idealized notions of justice -- assuming the operationalization to be a technical question, not a normative one.

In the context of our paper, we thus differentiate:
\begin{itemize}
    \item \textbf{Theory of distributive justice:} Ideal distribution of benefits/harms among individuals
    \item \textbf{Structural injustices:} Unfair distribution of deviations from the perfect fulfillment of the theory of distributive justice among groups
\end{itemize}

Notice that for both the theory of distributive justice and structural injustices, we have to define \textit{how} something should be distributed. The difference is in \textit{what} is distributed: For theories of distributive justice, it is \textit{benefits/harms themselves} that are distributed by a decision-making system. For structural injustices, it is the \textit{deviations} from the ideal distribution of benefits/harms. In both cases, we have to define what the ideal distribution should look like -- we could, e.g., ask for equality in benefits/harms or deviations. This is what we refer to as a \textbf{distributive pattern}. A distributive pattern does not say anything about what should be distributed or among whom it should be distributed -- it is, as the name suggests, just a pattern for how something (let us call this $X$) should be distributed among individuals/groups/etc. (let us call these $P$). The distributive pattern can, e.g., be egalitarian (i.e., demanding equality of $X$ among $P$), prioritarian (i.e., demanding the maximization of the sum of $X$ across $P$), sufficientarian (i.e., demanding that the $X$ of everyone in $P$ reaches a certain threshold $t$) or maximin (i.e., demanding the maximization of the minimal $X$ among $P$). Note that these are -- confusingly -- also the names of theories of distributive justice. The difference is that the respective theories of distributive justice do say something about what $P$ is (typically individuals) and what $X$ is.\footnote{Although there are typically many variations of a theory of distributive justice. There are, e.g., different variations of egalitarianism that demand equality in resources, well-being or how we relate to people.}
Distributive patterns are thus used by both theories of distributive justice as well as structural injustices. Note that when we derive fairness criteria to test for structural injustices under different theories of distributive justice in Section~\ref{sec:individual-group}, we will focus on the egalitarian pattern to measure structural injustices. However, Section~\ref{sec:mittelstadt} will discuss a possible alternative using the sufficientarian pattern.

\subsection{Terminology: Criteria and Metrics}

Before we dig in deeper, let us highlight the way in which we will use the terms \textit{criterion} and \textit{metric} in this paper, which are often used interchangeably in the literature.
In our paper, we view \textit{criteria} as binary conditions that can either be fulfilled or not.
If a criterion is not perfectly fulfilled (which is to be expected), it is helpful to be able to quantify the deviation from this perfect state. This is what we will refer to as a \textit{metric}. For example, if we expect equality in some \textit{measure} $M$ (i.e., we have an egalitarian criterion $\forall i,j \in P: M_i=M_j$), a possible metric is the absolute difference between the maximum and minimum values of the measures: $\max_{i \in P}(M_i) - \min_{j \in P}(M_j)$. Note that we can use a metric to create an \textit{approximate criterion} by demanding that the criterion is sufficiently well fulfilled as measured by, e.g., the metric falling below a certain threshold $t$, e.g., $\max_{i \in P}(M_i) - \min_{j \in P}(M_j) < t$.
An issue is that when we derive a metric for a criterion (or the other way around), we always have multiple options for doing so. These translations emphasize different aspects of what the criterion (or metric) measures, which corresponds to subtle variations in our understanding of the criterion (or metric).

Note that we will use the terms criterion and metric both in the context of distributive justice, where we will call them \textit{distributive justice criterion / metric}, and in the context of structural injustices, where we will call them \textit{fairness criterion / metric}. For the latter, we note that a more consistent term for our paper would be "structural justice criterion / metric", which we do not use as "fairness criterion / metric" is the established term in the algorithmic fairness literature.
The following list gives an overview of this terminology:

\begin{itemize}
    \item \textbf{Distributive justice criterion:} Measured at the individual level; tests whether the theory of distributive justice is perfectly fulfilled, which is presumably never the case in the real world
    \item \textbf{Distributive justice metric:} Measured at the individual level; measures deviation from perfect fulfillment of the theory of distributive justice
    \item \textbf{Fairness criterion:} Can be measured at different levels, but group fairness measures this at the level of socio-demographic groups; tests whether there are any signs of structural injustices, which we will probably always see in the real world as perfection is impossible to achieve, so while imperfect fulfillment of distributive justice tells us to look for signs of structural injustices, this would not be the way to do it in practice.
    \item \textbf{Fairness metric:} Group fairness again measures this at the level of socio-demographic groups; measures deviation from perfect fulfillment of the fairness criterion
\end{itemize}

Note that we cannot define a distributive justice criterion for all theories of distributive justice and that we cannot always define just a single unambiguous distributive justice metric. For optimization-type theories (see Section~\ref{sec:retracing}, we cannot define a distributive justice criterion that does not take other feasible distributions into account -- only a distributive justice metric. For criterion-type theories, the distributive justice criterion can be defined, but the metric is not unambiguous, and there are multiple options for how to define it. We will revisit this point when operationalizing maximin in Section~\ref{sec:individual-group}.

\subsection{Implications for Algorithmic Fairness}

In current discussions of algorithmic fairness, this differentiation has not been made clear. Instead, the concepts of distributive justice, structural injustices and distributive patterns have been mixed. We argue that algorithmic fairness has so far only looked at structural injustices but has not named this as such and has not made clear how this is related to theories of distributive justice.

When we want to test whether there are structural injustices in a given distribution -- as we claim we do in algorithmic fairness -- we thus have to define the following: (1) What theory of distributive justice the decision-making system is supposed to follow (because what structural injustice means depends on this contextual information) and then (2) how we define structural injustice in this context. Checking for perfect fulfillment of fairness criteria is not a realistic option in practice, so practitioners need both a distributive justice metric and a fairness metric. Existing group fairness criteria apply an egalitarian distributive pattern to test for structural injustices. However, note that we are not assuming that structural injustice is necessarily defined as the the violation of an egalitarian pattern. We will deal with the question of how structural injustice could be measured with other patterns, e.g., in sufficientarian terms, in Section~\ref{sec:mittelstadt}.

\begin{figure}[h]
\centering
\includegraphics[width=0.45\textwidth]{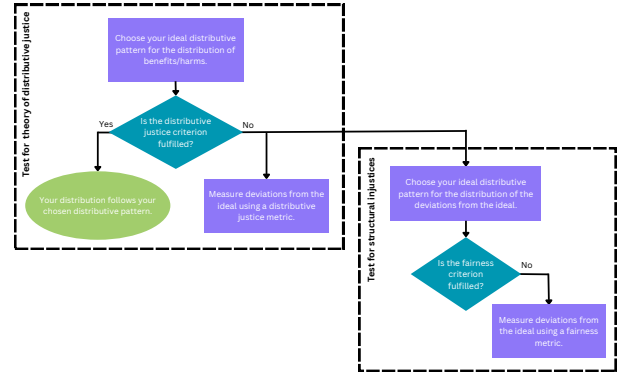}
\caption{Partial flow chart for checking the fulfillment of a theory of distributive justice and potential structural injustices in the deviations from that theory.}
\label{fig:flow-chart}
\end{figure}

Fig. \ref{fig:flow-chart} outlines the mapping of a theory of distributive justice and structural injustices. Note that only a part of the process of evaluating distributive justice and structural injustice is shown there as there is no obvious procedure for this when a distributive pattern is not perfectly fulfilled. Ideally, both the distributive justice metric and the fairness metric show small enough deviations from the ideal, but these metrics might also be at odds -- in which case trade-offs might be needed.

Ideally, we could demand for both the chosen distributive justice metric and fairness metric to be optimized. However, they might be at odds and the question of how to balance these two is a moral one. When we then have multiple possible models or distributions to choose from, it could make sense for practitioners to quantify the priority of the distributive justice metric compared to the fairness metric to help them choose a model if the better fulfillment of the theory of distributive justice and reduced structural injustices are at odds. One could, for example, imagine a Pareto front that shows the trade-offs between the two. One option would be to make the fairness criterion a constraint and then select the model that best fulfills the theory of distributive justice one has chosen.\footnote{This is what Rawls demands in his second principle~\cite{rawls1999theory}: As Rawls puts fair equality of opportunity as a condition over the difference principle, his approach would be to filter for all distributions that do not show signs of structural injustices (violations of the fair equality of opportunity principle) and to then select the one that best fulfills the difference principle. He calls this the lexical priority of fair equality of opportunity relative to the difference principle: “The second principle of justice is lexically prior to the principle of efficiency and to that of maximizing the sum of advantages; and fair opportunity is prior to the difference principle” \cite[p. 266]{rawls1999theory}.
}

Our takeaway from this is that whatever theory of distributive justice we believe our model should follow sets the basis for how we measure structural injustice: These two evaluations are not independent of each other. Rather, the measure of structural injustice depends on the chosen theory of distributive justice. The algorithmic fairness literature thus has to discuss fairness criteria / metrics in the context of their respective theory of distributive justice.

\section{From Theories of Distributive Justice to Fairness Criteria}\label{sec:individual-group}

In this section, we will show how the way in which we view distributive justice impacts algorithmic fairness. Specifically, we will go over three theories of distributive justice (egalitarianism, sufficientarianism, and maximin) and discuss how to check the fulfillment of the theory of distributive justice as well as potential structural injustices.

\subsection{Egalitarianism}\label{sec:op-egalitarianism}

We start with egalitarianism as this is the theory of distributive justice that group fairness criteria have retroactively been tied to (see Section~\ref{sec:retracing}).

\subsubsection{Test for Theory of Distributive Justice}

\paragraph{Distributive Justice Criterion}

If egalitarianism is perfectly fulfilled, everyone gets exactly the same. We can, therefore, evaluate egalitarianism as a theory of distributive justice at the individual level with the following justice criterion:

\begin{proposition}
The \textbf{distributive justice criterion for egalitarianism} is that all (equally deserving) individuals have the same utility: $\forall i,j \in P: u_i = u_j$.
\end{proposition}

\paragraph{Distributive Justice Metric}

If the justice criterion is not perfectly fulfilled, our rethinking of distributive justice has shown that we can check whether the distributive pattern is approximately fulfilled.
When we want to check whether our distribution fulfills egalitarianism well \textit{enough}, we could come up with a metric that measures the deviation from perfect fulfillment.

Examples of possible \textbf{distributive justice metrics} for egalitarianism are:
\begin{itemize}
    \item The Gini coefficient~\cite{gini1912variabilita} is a popular metric to measure inequality that is usually used to measure wealth or income inequality, e.g., in a country
    \item Variance in utilities across the entire population: $\text{Var}(U)$
    \item Difference between the maximum and minimum utility:\\ $\max_{i \in P} u_i - \min_{j \in P} u_j$
    \item Ratio of the maximum and minimum utility: $\dfrac{\max_{i \in P} u_i}{\min_{j \in P} u_j}$
\end{itemize}

We could then set a threshold for the metric to define how far the distribution can deviate from perfection to still be called "approximately just". This gives us an \textbf{approximate distributive justice criterion}.
The aim is to say "While not every utility is exactly equal, the utilities are similar enough according to our chosen metric."

\subsubsection{Test for Structural Injustices}

As we know, when the theory of distributive justice is not perfectly fulfilled, this gives rise to potential biases. This is what we would then consider to be a sign of structural injustice, which we have to check for separately.
Let us first figure out what structural injustices would look like in this context.

In the following, example 1 shows what structural injustice could look like for a distribution that approximately fulfills the theory of distributive justice. Example 2 then highlights that a distribution that does not show these signs of structural injustice does not have to be egalitarian though -- thereby showing that there is indeed a difference in checking for an approximate just distribution according to egalitarianism and checking for the absence of structural injustice.

\begin{itemize}
    \item \textbf{Example 1:} Assume that individuals are graded on an exam. We pick the students that are equally deserving of a B to check the grading process. With this background information, everyone in our set should get a B. Assume further that there are two groups, group 1 and group 2. The teacher is biased against group 2, which means they give every group 2 student a C but every group 1 student the B they deserve.\footnote{Note that it does not actually matter that they get grades close to B. They could all get a D and the test for equality / low variance would still pass since all that matters is that students get (about) the same grade.} Of course, the individual-level check for egalitarianism would fail as not every student gets the same grade. However, if we assume group 2 is small in size, the Cs might just be a few outliers that do not influence the variance much. We might thus say that the distribution is just. However, we cannot tell yet if there are structural injustices. For this, we have to check the grades of group 1 and group 2. We could, for example, compare grade averages to realize that the teacher systematically grades group 2 students lower than group 1 students. We thus need the fairness criterion to check for structural injustice.
    \item \textbf{Example 2:} Assume that instead of distributing bank loans based on an applicant's properties, a bank clerk just tosses a coin. Everyone has a 50-50 chance of getting the loan. While there is certainly something to be said about procedural justice here (a loan grant should probably not be decided based on a coin toss), let us focus on outcome fairness for now. Here, it seems that the resulting loans are not fair on an individual level if we think that getting a loan should at least be somewhat related to whether we can repay the loan or some other individual characteristics. If we focused on equally deserving individuals, say those who are able to repay the loan, we see that there is high variance, so our distribution would not be seen as just by the distributive justice metric. However, we also know for sure that whether you get a loan or not is not causally influenced by group membership -- there is thus no structural injustice, which is what we try to figure out with our metric. A fairness criterion comparing group averages would also show us that there is no systematic bias -- assuming that there are plenty of people asking for loans in both groups, so that the loan approval rates are similar in both groups.
\end{itemize}

These two examples show that testing for the theory of distributive justice and testing for structural injustices are two different things and that both are necessary. 

\paragraph{Fairness Criterion}

Fairness criteria are supposed to detect possible structural injustices, which we defined as unfairly distributed deviations from an ideal distribution.
We are thus looking for patterns of systematic differences between groups. This shows itself when the expected utilities of both groups are compared. A (notable) difference in these expectation values hints at possible structural injustices.

\begin{proposition}
A possible \textbf{fairness criterion for egalitarianism} is that the expected utilities of all groups are equal: $\forall a,b \in A: E(U|a)=E(U|b)$.
\end{proposition}

This is essentially how standard group fairness criteria operate.\footnote{Except that they do not take a utility-based approach and instead compare binary decisions or outcomes across groups.} This tells us that standard group fairness criteria seem to assume an egalitarianism as the theory of distributive justice and then also use the egalitarian distributive pattern to measure potential structural injustices.

\paragraph{Fairness Metric}

Since this fairness \textit{criterion} is almost impossible to perfectly fulfill in practice, we also want to measure the deviation from the criterion, which we formalize as \textit{metrics}.
The list of possible metrics is similar to the list of possible distributive justice metrics: the Gini coefficient, variance, a difference or ratio, etc. This similarity occurs because both criteria (the distributive justice criterion and the fairness criterion) demand equality in something (utilities/expected utility) across a set (of individuals/groups).

\subsection{Sufficientarianism}\label{sec:op-sufficientarianism}

We now apply our approach to sufficientarianism to derive appropriate fairness criteria and metrics from this. This will give us to an important insight: To check for structural injustices in a sufficientarian distribution, we still use an egalitarian pattern -- what changes compared to the fairness criteria for egalitarian distributions is the measure that we demand to be equal.

\subsubsection{Test for Theory of Distributive Justice}

\paragraph{Distributive Justice Criterion}

If sufficientarianism is perfectly fulfilled, everyone's utility is above the predefined minimum threshold:

\begin{proposition}
The \textbf{distributive justice criterion for sufficientarianism} is that all (equally deserving) individuals have a utility above the threshold $t$: $\forall i \in P: u_i > t$.
\end{proposition}

\paragraph{Distributive Justice Metric}

An intuitive distributive justice metric is the share of individuals with a utility above the threshold $t$. To get an approximate distributive justice criterion, this metric can be demanded to be sufficiently high.

\subsubsection{Test for Structural Injustices}

We again discuss two examples to show that an approximately just distribution does not have to go hand in hand with the absence of structural injustices to help us understand what would constitute structural injustices in the context of sufficientarianism.

\begin{itemize}
    \item \textbf{Example 1:} Picture a world in which 95\% of individuals reach the threshold, which is assumed to be sufficient for the distribution to be seen as approximately fulfilling sufficientarianism. However, the 5\% not reaching the threshold are all in group 1 while the 95\% reaching the threshold are in group 2. There is clearly an issue of structural injustice despite the approximate distributive justice test passing.
    \item \textbf{Example 2:} Imagine that approximately equal shares of group 1 and 2 are above the threshold, so that there is no issue of structural injustice. Now imagine, however, that overall only 1\% of the entire population is above said threshold. Sufficientarianism is then not reached even though there is no issue of structural injustice.
\end{itemize}

\paragraph{Fairness Criterion}\label{sec:sufficientarianism-fairness-criterion}

Sufficientarianism assumes that there are utility thresholds that, ideally, every individual should reach. This could, for example, be the case in the distribution of medicine where everyone needs one pill to become healthy again and half pills do not work. Then the threshold is one pill. It does not matter whether an individual receives more than one pill -- they will be healthy either way. Likewise, it does not matter whether they receive half a pill or no pill as they will stay sick. We need to keep this in mind when thinking about what structural injustices are as we first need to define what deviations from the ideal distribution are. Under the ideal distribution, everyone reaches the threshold. Deviations are cases where people do not reach the threshold. It does not matter how far they fall below the threshold (a half pill does not cure the sick patient), so for the deviations, we only care whether they are below the threshold; this is thus a binary measurement. Then, the question of structural injustice is: What is an unfair distribution of these deviations?
We propose the following fairness criterion:
\begin{proposition}
A possible \textbf{fairness criterion for sufficientarianism} is that the share of individuals with utilities above the threshold $t_u$ is equal across all groups: $\forall a,b \in A: E(U'|a)=E(U'|b)$, where $U'=I(U>t)$, so $U'$ models the resulting utility as binary: either one is above the threshold ($U'=1$) or below ($U'=0$).
\end{proposition}

Here, we notice that checking for structural injustices under sufficientarianism leads us to an egalitarian distributive pattern: To check if there is a systematic bias against a group, we check if groups are (approximately) treated equally. What changes compared to Section~\ref{sec:op-egalitarianism} is \textit{what} we demand to be equal: It is not $E(U)$ but $E(U')$. One can then define \textit{fairness metrics} and \textit{approximate fairness criteria} in the same manner as we did for the egalitarian theory of distributive justice in Section~\ref{sec:op-egalitarianism}.

\subsection{Maximin}

Let us now apply the same thinking to maximin. Remember that maximin is an optimization-type theory of distributive justice (see Section~\ref{sec:retracing}), which has an influence on the operationalization.

\subsubsection{Test for Theory of Distributive Justice}

\paragraph{Distributive Justice Metric}

If we follow maximin, the utility of the worst-off individual $min_{i \in P}(u_i)$ is maximized. However, assuming that decision-making systems are not perfect, we might not want to focus on individuals. To avoid looking at the single worst-off individual as there could always be an outlier, we propose measuring the expected utility of the worst-off 5\% of individuals.
Note that here we only derive a metric and not a criterion due to maximin being a optimization-type theory of distributive justice.\footnote{We could turn it into a criterion by asking that among all possible distributions $R$, we choose the one that maximizes our metric.}

\subsubsection{Test for Structural Injustices}

How does structural injustice play out for maximin? Imagine the following two distributions of wages per hour of work where all people whose names start with the letter "A" are in group A and all people whose names start with the letter "B" are in the marginalized group B. Assume that group A and B are roughly equal in their natural talents and motivations relevant to the job. Assume also that the distributions exhibited here are statistically significant, so we assume that the inequality is due to unequal opportunities.
\begin{itemize}
    \item Anna: 10€, Berta: 20€, Anton: 30€, Basti: 40€, Adriana: 50€, Barbara: 60€
    \item Berta: 10€, Basti: 20€, Barbara: 30€, Anna: 40€, Anton: 50€, Adriana: 60€
\end{itemize}

Both distributions are equally good according to maximin as the worst-off person always has 10€ (and using the approximate version, the worst-off 5\% are also equal in their utility). Given that maximin cares about the worst-off individual, this theory of justice assigns moral value to the one that is worst-off. This gives us reason to believe that someone who subscribes to maximin and cares about structural injustices should find the first distribution preferable to the second one.\footnote{Note that this cannot be avoided by demanding a leximin distribution as the two distributions are also equally good according to leximin. Yet, since leximin cares about those who are worst off, someone who cares about structural injustices should again prefer the first distribution.} There are multiple possible explanations as to why they should prefer the first distribution:
\begin{itemize}
    \item The worst-off individuals of the second distribution are all in group B.
    \item The worst-off individuals of group B (Berta, Basti) are worse off than the worst-off individuals of group A (Anna, Anton) in the second distribution.
\end{itemize}

To evaluate structural injustices, we do not just check what group the single worst-off individual belongs to: of course, this individual has to belong to some group, and this does not yet constitute a structural injustice against that group. We argue that it only becomes problematic once we see a pattern at the group level. We, therefore, look at the 5\% worst-off individuals (in the population/per group) -- similar to the distributive justice metric.

\paragraph{Fairness Criterion}

From this, we could come up with different fairness criteria:
\begin{itemize}
    \item Fairness demands that it is (approximately) equally likely for members of each group to be in the worst-off 5\%.
    \item Fairness demands that the expected utility of the worst-off 5\% in each group have an (approximately) equal expected utility.
\end{itemize}

Again, these fairness criteria only tell us something about fairness and not about which distribution is more just according to maximin. Imagine, for example, a distribution where half of the worst-off 5\% are in group A and the other half in group B -- or a distribution where the worst-off 5\% of group A have the same expected utility as the worst-off 5\% of group B. A second distribution does not fulfill these, but the worst-off individual is (or the worst-off 5\% are) better off in distribution 2. In that case our fairness criterion would tell us to prefer distribution 1, but maximin would actually require distribution 2. This again shows that the criteria and metrics that we use to evaluate the justness of a distribution are different from the criteria and metrics that we use to evaluate structural injustices.

Analogous to fairness criteria for an egalitarian (Section~\ref{sec:op-egalitarianism}) and a sufficientarian (Section~\ref{sec:op-sufficientarianism}) theory of distributive justice, we can turn the fairness criteria for maximin into \textbf{fairness metrics} and \textbf{approximate fairness criteria}. Note that by picking two options to translate maximin to a fairness criterion, we get a multitude of fairness metrics as there are also multiple options for translating each of these fairness criteria into a metric. What does this tell us and how should we choose between them? We argue that they highlight different aspects of what matters about maximin and structural injustices. Depending on one's views on this, one can choose different paths and end up with a different criterion or metric.

\section{Comparison to Naive Operationalizations}\label{sec:naive-operationalizations}

As previously discussed, examples to illustrate theories of distributive justice are often given at the individual level. However, group fairness criteria are defined at the group level. Our approach shows fairness criteria depend on the theory of distributive justice they were created for. This section will highlight the necessity of our approach by comparing the resulting fairness criteria to a more naive approach that simply takes the theory of justice defined at the individual level and replaces the term "individual" with "group" and "utility" with "expected utility".
Such an approach would result in the following fairness criteria:

\begin{itemize}
    \item \textbf{Egalitarianism:} Every individual has the same utility. $\rightarrow$ Every group has the same expected utility.
    \item \textbf{Sufficientarianism:} Every individual's utility is above the threshold. $\rightarrow$ Every group's expected utility is above the threshold.
    \item \textbf{Maximin:} Maximize the utility of the worst-off individual. $\rightarrow$ Maximize the expected utility of the worst-off group.
\end{itemize}

These naive operationalizations of theories of justice as group-level fairness criteria are a tempting way to test for structural injustice -- after all, Section~\ref{sec:op-egalitarianism} showed that it seems to work for egalitarianism. In this section, we will discuss why the naive operationalization of theories of justice as group-level fairness criteria can turn out to be logically inconsistent with said theory of distributive justice and mismatch our intuitions about said theory.

\subsection{Sufficientarianism}\label{sec:naive-sufficientarianism}

\begin{figure}[h]
\centering
\includegraphics[width=0.3\textwidth]{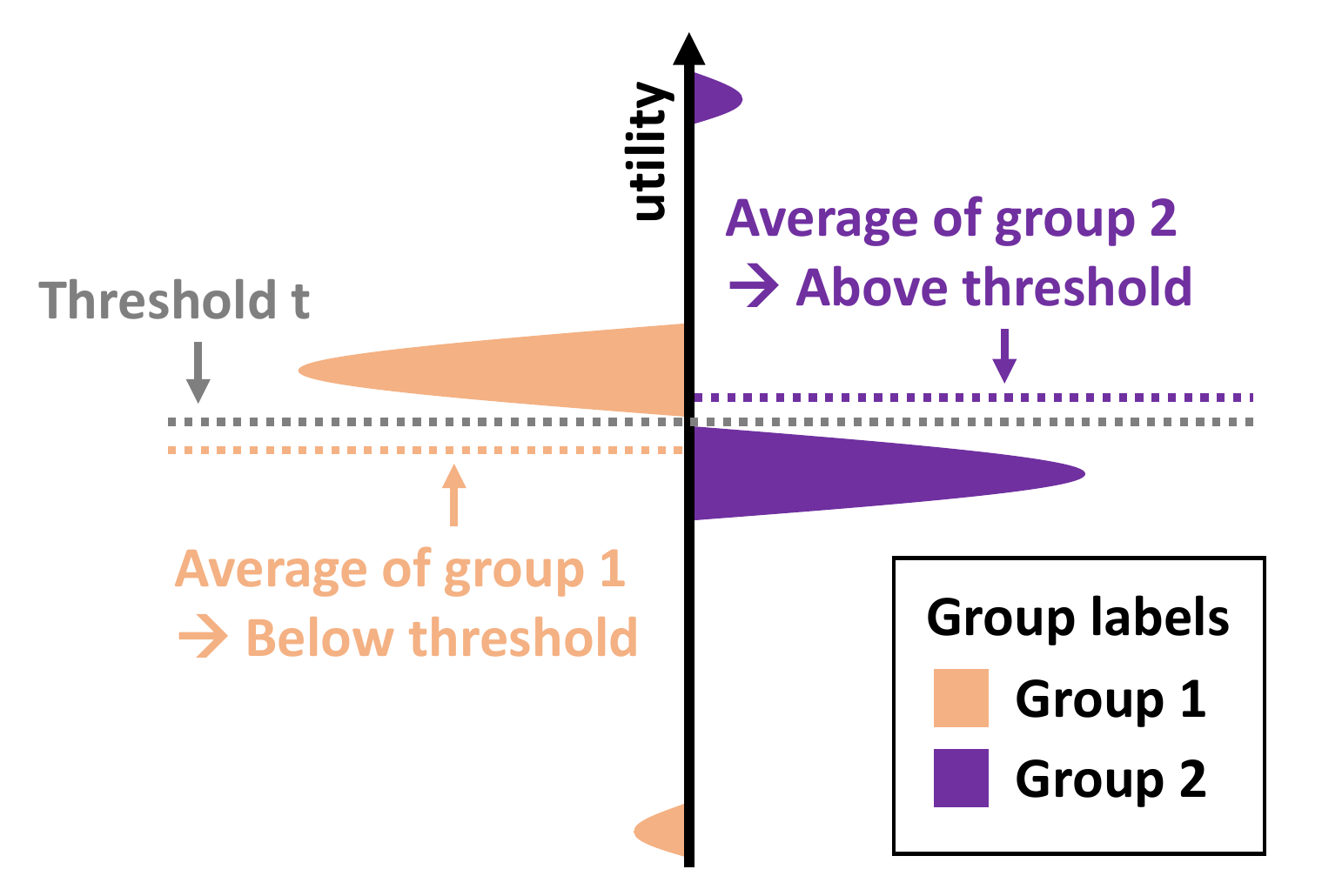}
\caption{Illustration of the naive operationalization of sufficientarianism}
\label{fig:sufficientarianism-counterexample}
\end{figure}

Our approach to operationalizing sufficientarianism results in a different fairness criterion than the naive operationalization.\footnote{One might be tempted to think that \citet{mittelstadt2023unfairness} suggests such a naive operationalization of sufficientarianism: instead of demanding that all individuals reach a certain threshold,~\cite{mittelstadt2023unfairness}'s fairness criterion demands that all groups reach a certain threshold.
However, we argue that~\cite{mittelstadt2023unfairness} should actually be understood differently. Section~\ref{sec:mittelstadt} will therefore discuss a more charitable interpretation of \citet{mittelstadt2023unfairness}'s work as well as how their work fits into our understanding of distributive justice.}
As we will show in this section, the naive operationalization is not logically coherent with the idea of sufficientarianism.
To see this, imagine two groups. In group 1, 95\% of the members are above the threshold. In group 2, only 5\% are above the threshold. This is illustrated in Fig.~\ref{fig:sufficientarianism-counterexample}. Now imagine further that the 95\% above the threshold in group 1 are just barely above the threshold while the 5\% below the threshold are far below the threshold. The resulting average group utility is below the threshold, so the naive operationalization of sufficientarianism would constitute that group 1 does not fulfill the sufficientarian requirement. In group 2, however, the 95\% below the threshold are barely below the threshold while the 5\% above the threshold are far above the threshold. Here, the resulting average utility is above the threshold, meaning the naive operationalization would see sufficientarianism to be fulfilled -- despite 95\% of the group not reaching the threshold. Assuming that the threshold represents what is needed to survive, this naive operationalization would assume that group 2 is better off than group 1 even though 95\% of individuals in group 2 are starving while it is "only" 5\% in group 1 that are starving. This means that this naive operationalization does not tell us anything about structural injustices. Clearly, group 1 is better off according to sufficientarianism and group 2 is the disadvantaged group. Our fairness criterion, which checks if approximately equal shares from all groups are above the threshold, accounts for this. Moreover, this naive operationalization is not even fit to check if sufficientarianism is fulfilled. This is where our distributive justice metric is necessary, which checks if a large enough share of the population reaches the threshold.

\subsection{Maximin}

\begin{figure}[h]
\centering
\includegraphics[width=0.3\textwidth]{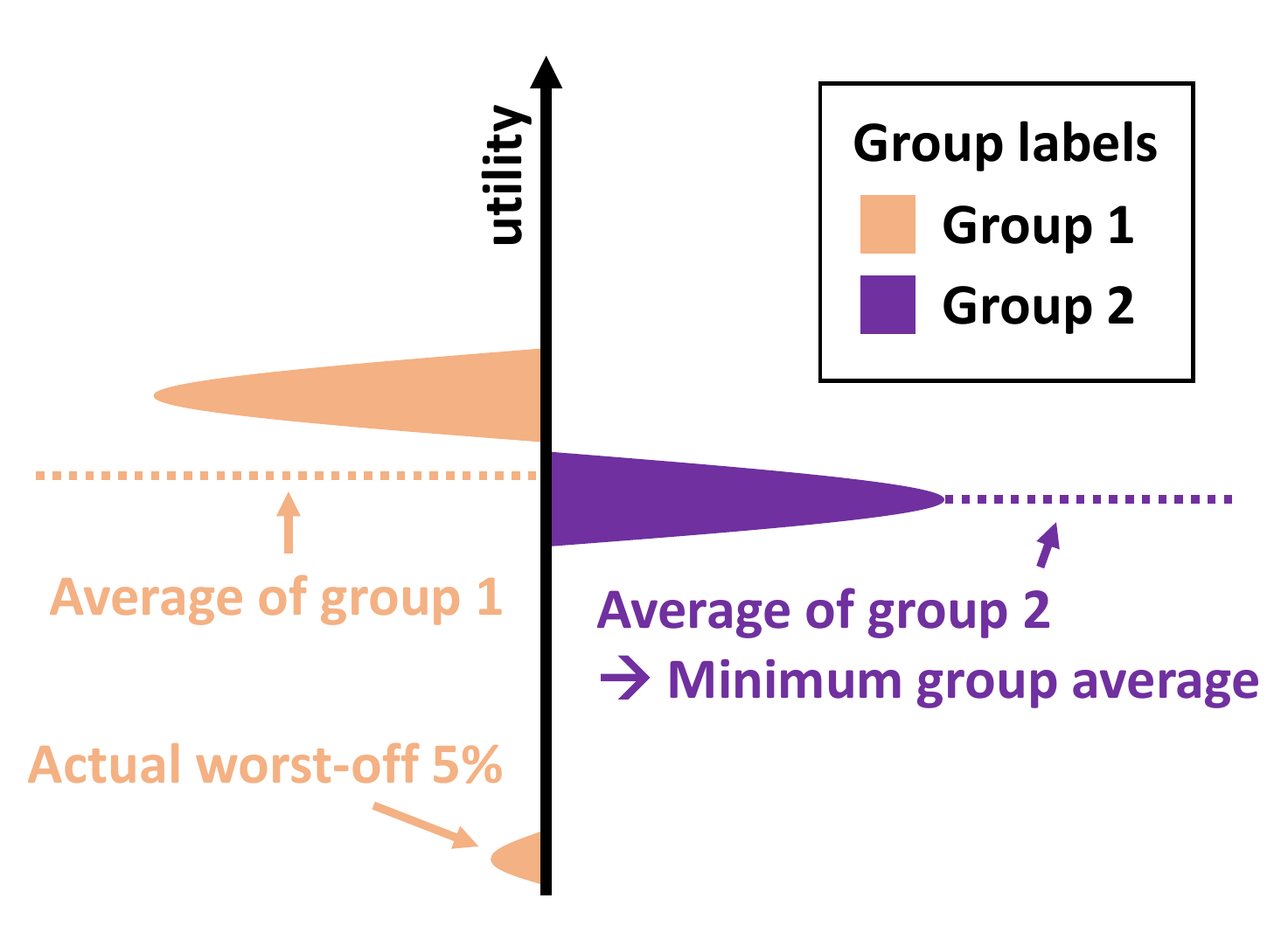}
\caption{Illustration of the naive operationalization of maximin}
\label{fig:maximin-counterexample}
\end{figure}

Both~\cite{heidari2019moral}'s\footnote{\cite{heidari2019moral} briefly discusses maximin as an alternative to egalitarianism citing the leveling down objection. They suggest that "the max-min distribution deviates from equality only when this makes the worst off group better off", but do not discuss why maximin can or should be analyzed at the group level.} and~\cite{martinez2020minimax}'s\footnote{\cite{martinez2020minimax} integrates maximin into a formalized fairness criterion, which has already been used in other influential papers such as~\cite{diana2021minimax}. To be precise, they actually use minimax, which is just the counterpart of maximin for situations in which what we compare across groups is something negative or harmful that groups want to minimize. The idea is thus to look at the group with the maximum value and compare this maximum value across possible distributions/models to then choose the minimum. To make it easier to follow~\cite{martinez2020minimax} in the context of our paper where we use maximin, we will adjust the following explanation of~\cite{martinez2020minimax} to make it maximin:  \cite{martinez2020minimax} suggests a multi-objective optimization that finds the distribution where the group with the lowest accuracy is maximized under the condition that this distribution is also Pareto-efficient across all groups' accuracy scores. This means that one would not select a distribution where the accuracy score of any group could be improved without worsening the accuracy score of the worst-off group. Fairness is thus measured as the worst accuracy score across all groups. The paper does not give a moral justification for why they care about accuracy scores.} maximin fairness metric correspond to the naive version of translating maximin to the socio-demographic group setting that we described at the beginning of Section~\ref{sec:individual-group}. The issue with this operationalization is that if the worst-off individuals are in a group with a lot of very well-off individuals, then the actual worst-off individuals are ignored in the distribution process. Imagine, for example, that the 5\% worst-off individuals are distributed across groups that are, on average, fairly well-off as seen for group 1 in Fig.~\ref{fig:maximin-counterexample}. Now imagine that another group (group 2 in Fig.~\ref{fig:maximin-counterexample}) has a lower average utility, but the worst-off individuals from that group are doing much better than the worst-off individuals from group 1. The worst-off individuals from group 1 are then ignored as we only maximize the utility of the group that is, on average, worst-off, so group 2. As the previous discussion of how to evaluate structural injustice for maxmin showed, there are other fairness criteria that are -- so we argue -- more in the spirit of maximin.

\section{Is Group Fairness Inherently Egalitarian?}\label{sec:mittelstadt}

In Section~\ref{sec:individual-group}, we defined fairness criteria under different theories of distributive justice.
This helps us check whether a sufficientarian or maximin decision-making system treats people of a disadvantaged group unfairly.
You may have noticed that for sufficientarianism, we demanded \textit{equality} in the share of groups' members above the threshold $t$ while for maximin, we proposed two fairness criteria that both demand \textit{equality} in a measure between groups.
One might ask if there are alternatives to this egalitarian fairness approach -- in particular because this type of egalitarianism still carries the risk of leveling down: When such a fairness criterion is enforced, it could actually happen that the share of members above the threshold is lowered in all groups (for sufficientarianism) or that, e.g., the expected utility of the worst-off individuals is lowered (for maximin).
\cite{mittelstadt2023unfairness} have therefore suggested another approach, which does not check for equality but instead demands some measure to reach a certain level.
They call this type of fairness constraint "leveling up" and describe it as follows:
"[If] we believe that people are being harmed by low selection rates, precision, or recall, instead of enforcing that these properties be equalised across groups, we can instead require that every group has, at least, a minimal selection rate, precision, or recall." \cite[p. 37]{mittelstadt2023unfairness}

As mentioned in Section~\ref{sec:naive-sufficientarianism}, this is reminiscent of sufficientarianism.
However, \citet{mittelstadt2023unfairness} never portray their proposed fairness criterion as an operationalization of sufficientarianism. Instead, they portray it in a way that is analogous to the way egalitarianism is used in group fairness criteria: Not as a criterion to check if this theory of distributive justice is fulfilled but rather as a fairness criterion to check if there are structural injustices at the group level.
In fact, "leveling up" does not represent a single fairness criterion but contains multiple possible fairness criteria:
Leveling up does not specify \textit{what} ought to be above the minimum threshold\footnote{They mention multiple options such as the true positive rate, the false positive rate, etc. While their examples focus on comparing binary decisions across groups ("The family of post-processing methods we consider tune a separate offset for each group, which alters the proportion of individuals in each group that receive a positive decision." \cite[p. 37]{mittelstadt2023unfairness}), they do not explicitly discuss their approach as being limited to binary decisions. We, therefore, interpret their proposal as being extendable to compare utility values.} -- just how egalitarianism does not specify \textit{what} ought to be equal (the \textit{equality of what} debate~\cite{sen1980equality}). Therefore, leveling up could ask for, e.g., the selection rates to be above a threshold $t_{\text{leveling-up}}$ ($\forall a \in A: P(D=1|A=a) > t_{\text{leveling-up}}$) or for the true positive rate to be above the threshold ($\forall a \in A: P(D=1|Y=1, A=a) > t_{\text{leveling-up}}$) -- just like egalitarianism contains multiple fairness criteria depending on what measure $M$ is chosen. A formalization of this making use of the generic measure $M$ and the threshold $t_{\text{leveling-up}}$ is $\forall a \in A: M(A=a) > t_{\text{leveling-up}}$.

We can combine our operationalization of fairness criteria for egalitarianism, sufficientarianism and maximin with~\citet{mittelstadt2023unfairness}'s leveling up. For an egalitarian distribution, we can demand that the expected utilities of groups reach a minimum threshold $t_{\text{leveling-up}}$: $\forall a \in A: E(U|A=a) > t_{\text{leveling-up}}$. Instead of demanding equal expected utilities, we would demand a minimum level for the expected utility of each group.
In the case of sufficientarianism, we can demand the share of individuals with utilities above the threshold $t$ to reach a certain threshold $t_{\text{leveling-up}}$ for all groups: $\forall a \in A: E(U>t|A=a) > t_{\text{leveling-up}}$ instead of demanding the share of individuals above the threshold $t$ to be equal across groups ($E(U>t|A=a)=E(U>t|A=b)$). This means that while we want equal shares of members from all groups to reach the threshold, we can now forgo equality if it harms a/all group and instead just demand that groups bring at least a certain share $t_{\text{leveling-up}}$ of individuals above the required threshold $t$. This notion is useful if you believe that it is, e.g., sufficient to get 90\% of individuals above the threshold in every group and it does not matter if one group gets 100\% of their members above the threshold while another group only gets the required 90\% above the threshold. If you believe that this is a sign of structural injustice, however, you should check for equal shares and thus use the egalitarian distributive pattern to test for structural injustices.
Similarly, we can adapt our fairness criteria for maximin. For the second fairness criterion (the expected utility of the worst-off 5\% in each group should be (approximately) equal across groups), we could, for example, demand that the expected utility of the worst-off 5\% in each group reaches a minimum threshold $t_{\text{leveling-up}}$.

There is thus a question of what we mean by structural injustice: Do we speak of structural injustice when the harms of deviating from a distribution that fulfills our chosen theory of distributive justice are unequally distributed? This would apply an egalitarian approach to structural injustice. Do we not actually care about equality in the distribution of deviations and instead care about not having too strong of a harmful deviation for each group? Then, we might want to take a sufficientarian approach to structural injustice (i.e., \citet{mittelstadt2023unfairness}'s "leveling up"). When choosing a group-level fairness criterion or metric, we thus first have to decide on what theory of distributive justice we find most appropriate. Then, we need to decide what we view as structural injustices under the chosen type of distribution.\footnote{Rawls's second principle~\cite{rawls1999theory}, for example, tells us that he views group-level fairness for maximin as egalitarian -- represented by what he calls "fair equality of opportunity".}
However, one could reasonably disagree on this question as it does not have a straightforward answer and depends on how one views what constitutes structural injustices.

\section{Conclusion}\label{sec:conclusion}

Group fairness criteria are typically described as "egalitarian". A lot of the algorithmic fairness literature, therefore, seems to assume that fairness criteria are the operationalization of theories of distributive justice. However, we argue that this is not true: While theories of distributive justice demand a certain distribution of benefits/harms among individuals, we claim that fairness criteria actually test whether the inevitable deviations from said ideal are unfair to certain groups, i.e., systematically biased against them. This test for a systematic bias in the deviations often uses an egalitarian distributive pattern and is therefore confused with the egalitarian theory of distributive justice. We show how this refined conceptual thinking affects what fairness criteria we use when we have a distribution that is supposed to follow theories of distributive justice other than egalitarianism.

\paragraph{Implications}
In current discussions of algorithmic fairness, the concepts of theories of distributive justice, structural injustices and distributive patterns are often confused.
Yet, recognizing these concepts and their relationships is crucial as the measurement of structural injustices depends on the chosen theory of distributive justice. When testing for structural injustices, this means first defining the theory of distributive justice and also testing whether it is fulfilled. When proposing new fairness criteria, this means discussing the theory of distributive justice in the context in which these fairness criteria can be used.
Our paper is thus a call for the algorithmic fairness literature to clearly differentiate between theories of distributive justice and structural injustices.
Future work should look into the subtle normative differences between the multiple fairness metrics that can be derived from a single fairness criterion and the other way around. This could provide valuable guidance to practitioners. It would also be interesting to apply the demonstrated approach to derive fairness criteria for other theories of distributive justice such as prioritarianism.

\begin{acks}
Thank you to Joachim Baumann for his valuable feedback and input on several versions of this draft. We also want to thank Marcello Di Bello for his thoughtful feedback on this paper.
This work was supported by the National Research Programme “Digital Transformation” (NRP 77) of the Swiss National Science Foundation (SNSF), grant number 187473.
\end{acks}

\bibliographystyle{ACM-Reference-Format}
\bibliography{main}


\end{document}